\documentclass[a4paper,11pt]{article}
\usepackage{aaskaiid}
\usepackage{orcidlink} 
\usepackage{wrapfig}
\newcommand*\degr{\ensuremath{^\circ}}
\newcommand*\arcmin{\ensuremath{^\prime}}

\title{A 3D tomography of the Local Bubble with SKA-Low}
\ShortTitle{3D tomography with SKA Low}

\author[1]{Xiaohui Sun\, \orcidlink{0000-0002-3464-5128}}
\ShortName{Sun et al.}  
\author[2,3]{George Heald\, \orcidlink{0000-0002-2155-6054}}
\author[4,5,6]{Andrea Bracco\,\orcidlink{0000-0003-0932-3140}}

\affiliation[1]{School of Physics and Astronomy, Yunnan University, Kunming 650500, China}
\emailAdd{xhsun@ynu.edu.cn}
\affiliation[2]{SKA Observatory, SKA-Low Science Operations Centre, 26 Dick Perry Avenue, Kensington WA 6151, Australia}
\affiliation[3]{CSIRO, Space \&\ Astronomy, 26 Dick Perry Ave, Kensington WA 6151, Australia}
\emailAdd{George.Heald@skao.int}
\affiliation[4]{LUX, Observatoire de Paris, Université PSL, Sorbonne Université, CNRS, 75014 Paris, France}
\affiliation[5]{INAF–Osservatorio Astrofisico di Arcetri, Largo E. Fermi 5, 50125 Firenze, Italy}
\affiliation[6]{Laboratoire de Physique de l'Ecole Normale Sup\'erieure, ENS, Universit\'e PSL, CNRS, Sorbonne Universit\'e, Universit\'e de Paris, 75005 Paris, France}
\emailAdd{andrea.bracco@obspm.fr}

\abstract{We know we reside in the Local Bubble (LB), but we know little about the magnetized gas inside the LB. The SKA-Low with more than half of the total number of stations distributed within 1~km distance provides enough short baselines and thus great surface brightness sensitivity. An SKA-Low polarization survey covering the frequency range of 50-350~MHz will deliver high-sensitivity and high-resolution images of diffuse polarized emission that originates from the local interstellar medium (ISM) and is probably related with the LB. Such a survey will also determine precisely the rotation measures (RMs) for the polarized structures using RM synthesis. These will allow us to reveal the 3D structure of the magnetized medium in the LB and to understand how the LB forms and evolves.}

\begin{document}
\newcommand{\actaa}{Acta Astron.} 
\newcommand{\araa}{ARA\&A} 
\newcommand{\aar}{A\&ARv} 
\newcommand{\aapr}{A\&ARv} 
\newcommand{\ab}{Astrobiol.} 
\newcommand{\aj}{AJ} 
\newcommand{\apj}{ApJ} 
\newcommand{\apjl}{ApJL} 
\newcommand{\apjs}{ApJSS} 
\newcommand{\ao}{Appl. Opt.} 
\newcommand{\apss}{Astro. \& Space Sci.} 
\newcommand{\aap}{A\&A} 
\newcommand{\aaps}{A\&AS.} 
\newcommand{\baas}{Bull. Am. Astron. Soc.} 
\newcommand{\caa}{Chinese A\&A} 
\newcommand{\cjaa}{Chinese J. A\&A} 
\newcommand{\cqg}{Class. Quantum Gravity} 
\newcommand{\gal}{Galaxies} 
\newcommand{\gca}{Geo. Cosmo. Acta} 
\newcommand{\icarus}{Icarus} 
\newcommand{\jcap}{JCAP} 
\newcommand{\jgr}{J. Geophys. Res.} 
\newcommand{\jgrp}{J. Geophys. Res. Planets} 
\newcommand{\jqsrt}{J. Quant. Spectrosc. Radiat. Transf.} 
\newcommand{\memsai}{Mem. SAIt} 
\newcommand{\mnras}{MNRAS} 
\newcommand{\nat}{Nature} 
\newcommand{\nastro}{Nat. Astron.} 
\newcommand{\ncomms}{Nat. Commun.} 
\newcommand{\nphys}{Nat. Phys.} 
\newcommand{\na}{New Astron.} 
\newcommand{\nar}{New Astron. Rev.} 
\newcommand{\physrep}{Phys. Rep.} 
\newcommand{\pra}{Phys. Rev. A} 
\newcommand{\prb}{Phys. Rev. B} 
\newcommand{\prc}{Phys. Rev. C} 
\newcommand{\prd}{Phys. Rev. D} 
\newcommand{\pre}{Phys. Rev. E} 
\newcommand{\prx}{Phys. Rev. X} 
\newcommand{\prl}{Phys. Rev. Let.} 
\newcommand{\psj}{Planet. Sci. J.} 
\newcommand{\planss}{Planet. Space Sci.} 
\newcommand{\pnas}{Proc. Natl Acad. Sci. USA} 
\newcommand{\procspie}{Proc. SPIE} 
\newcommand{\pasa}{PASA} 
\newcommand{\pasj}{PASJ} 
\newcommand{\pasp}{PASP} 
\newcommand{\rmxaa}{RMXAA} 
\newcommand{\sci}{Science} 
\newcommand{\sciadv}{Sci. Adv.} 
\newcommand{\solphys}{Sol. Phys.} 
\newcommand{\sovast}{Soviet Ast.} 
\newcommand{\ssr}{Space Sci. Rev.} 
\newcommand{\uni}{Universe} 

\setlength{\bibsep}{0.0pt}  
\maketitle

\section{Introduction}

The polarization sky of Galactic diffuse emission at low frequency ($\lesssim300$~MHz) has only been explored recently owing to high-sensitivity telescopes such as the Murchison Widefield Array (MWA) and the Low-Frequency Array (LOFAR). 

The low frequency emission is from synchrotron radiation of relativistic electrons spiraling the magnetic field, which is linearly polarized. When a linearly polarized wave propagates in the magnetoionic interstellar medium, the polarization angle rotates due to an effect called Faraday rotation. The amount of rotation is proportional to the wavelength squared, and the coefficient is the rotation measure (RM). With broadband multichannel receiver backends at modern telescopes such as MWA, LOFAR, and SKA, RMs can be derived using RM synthesis~\citep{Burn+66, Brentjens+05}, which transfers the observed $\mathcal{P}(\lambda^2)\equiv Q(\lambda^2)+iU(\lambda^2)$ to $F(\phi)\equiv Q(\phi)+iU(\phi)$ using the Fourier transform, where $\lambda$ is the wavelength and $\phi$ is the Faraday depth defined as $\phi(\vec{r})=0.81\int_{\vec{r}}^{\rm observer}n_e B_\parallel{\rm d}l$. The integral is from the position inside the source $\vec{r}$ to the observer, $n_e$ is the electron density in cm$^{-3}$, $B_\parallel$ is the line-of-sight component of the magnetic field in $\mu$G, ${\rm d}l$ is the increment in path length in pc, and $\phi$ is in rad~m$^{-2}$. The Faraday spectrum~\citep[e.g.][]{Sun+15}, or the Faraday dispersion function~\citep{Brentjens+05}, $|F(\phi)|$, represents the emission structure of the source. RM is different from $\phi$. For a simple source without internal Faraday rotation, $|F(\phi)|$ approaches a $\delta$ function, the peak $\phi$ is equivalent to the RM of the source, and by searching for the location of the peak in $|F(\phi)|$, the RM can be derived. If the source is complex with multiple unresolved emission components, RM and $\phi$ are different; in this case, recovering the RM is challenging~\citep[e.g.][]{Sun+15}.

The advantage of polarization observations at low frequencies is that RM can be determined much more precisely than at high frequencies. Because $\mathcal{P}(\lambda^2)$ and $F(\phi)$ are Fourier transform pairs, the limited observation bandwidth means a sampling window in the $\lambda^2$ domain, which in turn causes a response function in the $\phi$ domain, called the RM spread function (RMSF). The full width at half magnitude (FWHM) of the RMSF, similar to the beam width in imaging, determines the resolution in the $\phi$ domain and thus the precision of the RM as $\sigma_{\rm RM}={\rm FWHM}/2{\rm SNR}$~\citep[e.g.][]{Vanderwoude+24}. Here, $\sigma_{\rm RM}$ is the RM uncertainty and $\rm SNR$ is the signal-to-noise ratio in polarized intensity. The FWHM is proportional to $1/\Delta\lambda^2$, where $\Delta\lambda^2$ is the maximum separation of $\lambda^2$. For a 50~MHz bandwidth centered at 200~MHz, falling in the frequency range of 50-350~MHz for SKA-Low, and an 810~MHz bandwidth centered at 1.31~GHz for the entire SKA-Mid Band 2, $1/\Delta\lambda^2$ is 1.46~m$^{-2}$ and 14.15~m$^{-2}$, respectively. This means that $\sigma_{\rm RM}$ at the lower frequency improves by an order of magnitude for the same SNR. If the central frequency shifts to the lower end of the SKA-Low frequency range or the bandwidth is wider, the improvement is even greater. 

On the other hand, there is also a disadvantage for observations at low frequencies, particularly for diffuse emission, which is depolarization. A detailed discussion on depolarization was presented by \citet{Sokoloff+98}. Here, we assess the depolarization factor, defined as the ratio of observed to intrinsic fractional polarization, at a typical frequency of 200~MHz. A small depolarization factor means a large depolarization. If the synchrotron-emitting and Faraday-rotating gasses are mixed, there is Faraday differential depolarization and the depolarization factor can be written as $\sin(2{\rm RM}\lambda^2)/2{\rm RM}\lambda^2$. For a small $|{\rm RM}|=5$~rad~m$^{-2}$, the depolarization factor is already 0.03, indicating a nearly complete depolarization. If a turbulent Faraday-rotating gas is present in front of a synchrotron-emitting gas, there is beam depolarization and the depolarization factor can be written as $\exp(-2\delta_{\rm RM}^2
\lambda^4)$, where $\delta_{\rm RM}$ is the scattering of RM in the foreground gas within the observation beam. For a small $\delta_{\rm RM}=0.5$~rad~m$^{-2}$, the depolarization factor is 0.08, also indicating a nearly complete depolarization. The depolarization becomes more severe at lower frequencies. 

Polarization observations of diffuse emission at low frequencies provide an opportunity to study the local interstellar medium (LISM) due to depolarization. Here, LISM refers to the ISM within several hundred parsecs around the Sun. The disadvantage now becomes an advantage. Provided a maximum $|{\rm RM}|$ of 5~rad~$^{-2}$, the maximum path length can be estimated as $L\approx |{\rm RM}| / 0.81<\nobreak n_e\nobreak ><\nobreak B_\parallel\nobreak >$, where $<\nobreak \ldots \nobreak >$ denotes an average. For $<\nobreak n_e \nobreak > \sim0.01$~cm$^{-3}$ from the thermal electron model YMW16~\citep{Yao+17} and $<\nobreak B_\parallel \nobreak >\sim2$~$\mu$G~\citep{Han+17}, $L$ is about 300~pc. RM scattering can be estimated as $\delta_{\rm RM}\approx 0.81<\nobreak n_e\nobreak >b\sqrt{Ll}/2\sqrt{3}$~\citep[e.g.][]{Gaensler+01}, where $b$ and $l$ are the strength and fluctuation scale of the random magnetic field, respectively. In the solar neighborhood, the strength of the random field is about 2-4 times that of the regular field~\citep[][]{Haverkorn+15}. To cause beam depolarization, the angular scale of the fluctuation should be smaller than the beam size $\Theta$, which means $l\lesssim L\Theta$. For $b\sim4$~$\mu$G, $\Theta\sim1\arcmin$, and $L\sim300$~pc, $\delta_{\rm RM}$ is estimated to be less than about 0.05~rad~m$^{-2}$, indicating that beam depolarization can be ignored. For a larger beam width or a longer path length, $\delta_{\rm RM}$ will be larger, causing a more severe beam depolarization. The maximum path length becomes smaller at lower frequencies. Therefore, polarized emission from within about 300~pc is observed at low frequencies, which probes the LISM. 

\begin{figure}[!htbp]
    \centering
	\includegraphics[width=0.98\columnwidth]{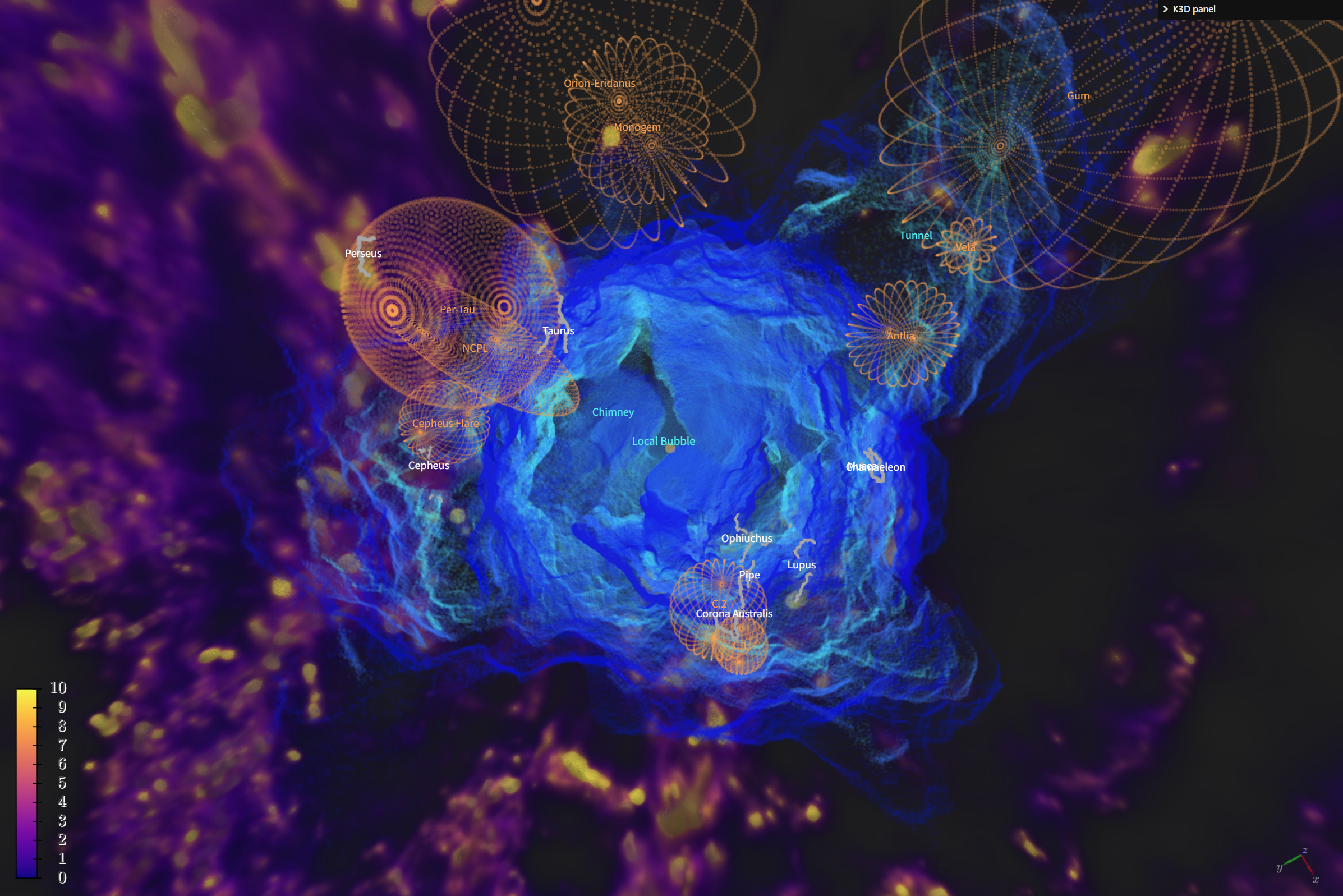}
    \caption{\it View of the local bubble (LB) in the context of solar neighborhood rendered from the 3D figure by \citet[][their Fig.~9]{O'Neill+24}. The bubble edges are outlined with blue color and the features are described in Table~2 of \citet{O'Neill+24}.}
    \label{fig:localbubble}
\end{figure}

Note that for the above estimate, we assume a slab of uniform statistical properties described by $\delta_{\rm RM}$. In reality, the ISM structure is highly non-uniformly distributed and the concept of a maximum path length, or polarization horizon, may be line-of-sight dependent. This is even more true with the possible 3D reconstruction of the cold gas in the first 1~kpc thanks to the latest dust maps~\citep[e.g.][]{Bracco+22}. 

The primary feature of the LISM is the local bubble (LB), where the Sun is roughly located at the center. The LB is a cavity with a very low density of hot gas that emits X-rays and is deficient in {\sc H\,i} and dust~\citep{Cox+87}. It is probably caused by 14-20 sequential supernovae in the last 10-15~Myr~\citep{Fuchs+06, Breitschwerdt+16, Zucker+22}. The LB can be best delineated in the dust maps that have been constructed recently. In Fig.~\ref{fig:localbubble}, we show a view rendered from the 3D figure by~\citet{O'Neill+24} based on the 3D dust map by~\citet{Edenhofer+24}. The blue color outlines the edges of the bubble, and the average distance from the edge to the Sun is about 170~pc. 

All-sky polarization observations at low frequencies will produce images of precise RMs and synchrotron emission contributed by the LISM, which will help to understand the origin and evolution of the LB. 

\section{Current Status}

\subsection{Modeling of the LB}

The simulations by \citet{Maconi+23} showed that the LB made a non-negligible contribution to dust polarization. \citet{Alves+18} showed that the LB can reproduce the dust polarization feature at 353~GHz within $30\degr$ of the Galactic poles. \citet{O'Neill+25} found that the LB has a non-negligible contribution to the dust and starlight polarization and derived a 3D magnetic field model of the LB.  

The LB also contributes to the RMs and the synchrotron emission. The synchrotron emissivity within about 1~kpc distance from the Sun was found to be much higher than that from beyond the distance based on observations of the absorption by {\sc H\,ii} regions~\citep{Roger+99, Nord+06}, suggesting a local excess of synchrotron emission. By including this local excess, the total intensity from the Galactic high latitude can be better reproduced~\citep{Sun+08, sun+10}. The cause of the local excess might be enhanced magnetic field or cosmic-ray electrons, which can be related to the LB. 

\citet{Korochkin+25} tried to model all-sky RMs and synchrotron emission at 22~GHz by including the LB in addition to the large-scale field model. They found that a model of the LB with a radius of 200~pc, a shell thickness of 30~pc, and an enhancement of magnetic field strength and thermal electron density by a factor of about 3 due to compressing led to better agreement with observations. This was corroborated by modeling RMs and synchrotron polarization emission at 30~GHz by~\citet{Pelgrims+25}. The RMs contributed by the LB are within $\pm10$~rad~m$^{-2}$ with a clear dependence on Galactic longitude, which can be examined with observations. The simulations by \citet{Maconi+25} produced frequency cubes of $Q$ and $U$ of diffuse emission at 1-5~GHz based on a model of the LB with enhanced thermal electron and cosmic-ray electron densities. By running RM synthesis, they found that RMs can be recovered from diffuse emission. Whether this holds for low frequencies needs further investigation. 

\subsection{Diffuse polarization sky at low frequencies}

The observations of diffuse polarized emission from the Galaxy have largely been driven by LOFAR and MWA using the RM synthesis technique. Low-frequency polarization with LOFAR, for instance, started observing ubiquitous polarization features at $|\rm RM|<20$~rad~m$^{-2}$, which have been associated with partially ionized gas, including {\sc H\,i} emission, at the wall of the LB~\citep{Bracco+20, Turic+21, Boulanger+24}.

For LOFAR, early observations of the 3C~196 field by \citet{Jelic+14} revealed coherent large-scale polarized structures. With the LOFAR two-meter sky survey~\citep[LoTSS,][]{Shimwell+17}, polarization images of about 10,\,000 deg$^2$ were obtained~\citep{VanEck+19, Erceg+22, Snidaric+23,Erceg+24a, Erceg+24}. The frequency range is 120-168~MHz divided into 480 channels each with a width of 97.6~kHz, corresponding to an FWHM of the RMSF of about 1~rad~m$^{-2}$. The resolution is about $5\arcmin$. The RMs of the polarization structures are mainly within $\pm10$~rad~m$^{-2}$ with a brightness temperature of a few K. Some of the structures are aligned with other tracers of the magnetic field, such as {\sc H\,i}, dust, and starlight polarization~\citep[e.g.][]{VanEck+17,Bracco+20}.   

\begin{figure}[!htbp]
    \centering
	\includegraphics[width=0.98\columnwidth]{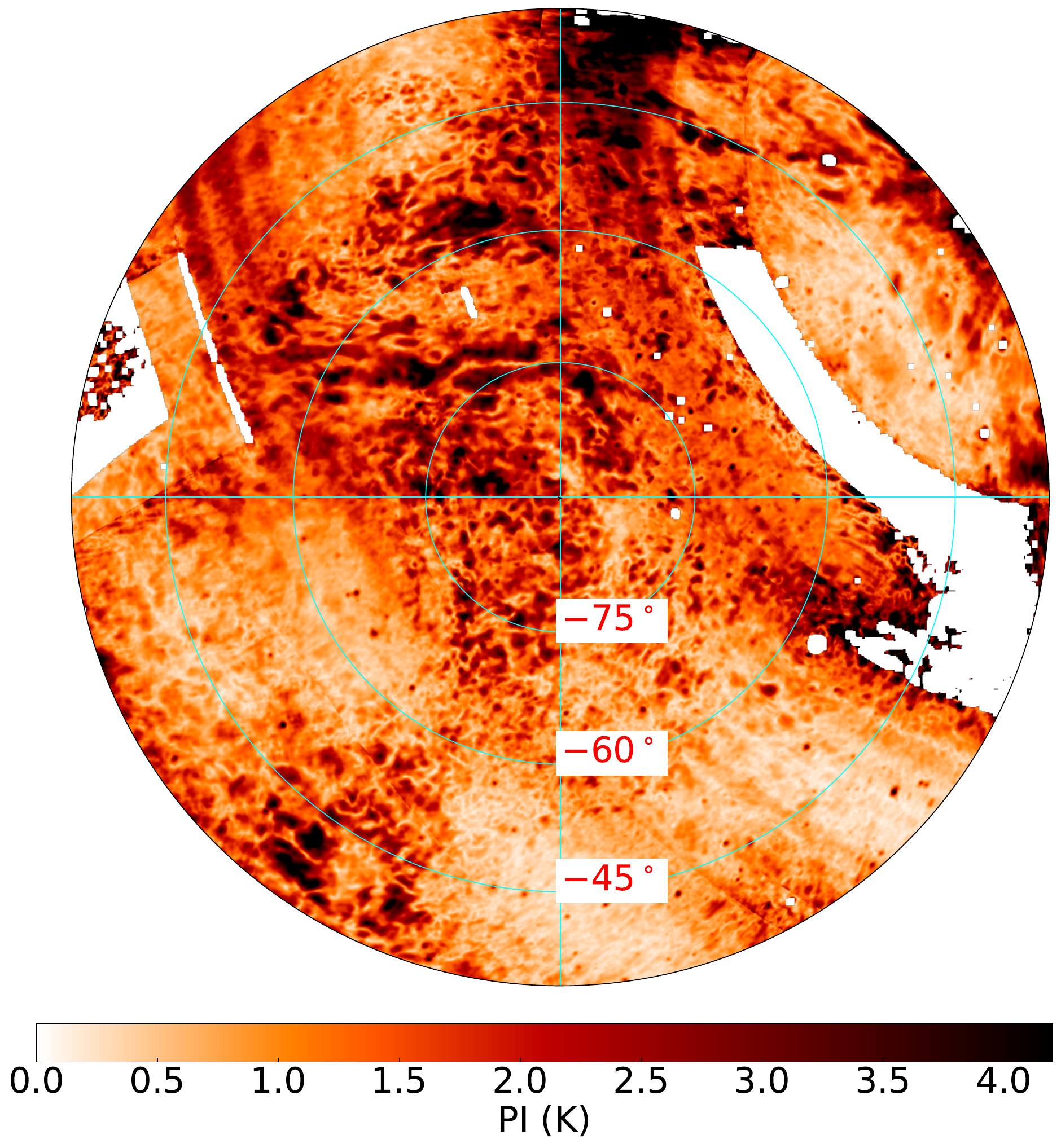}
    \caption{\it Preliminary image of polarized intensity at 215~MHz from MWA (Sun et al. in prep.). Areas with bad data or calibration are masked.}
    \label{fig:pi_mwa}
\end{figure}

\begin{figure}[!htbp]
    \centering
	\includegraphics[width=0.98\columnwidth]{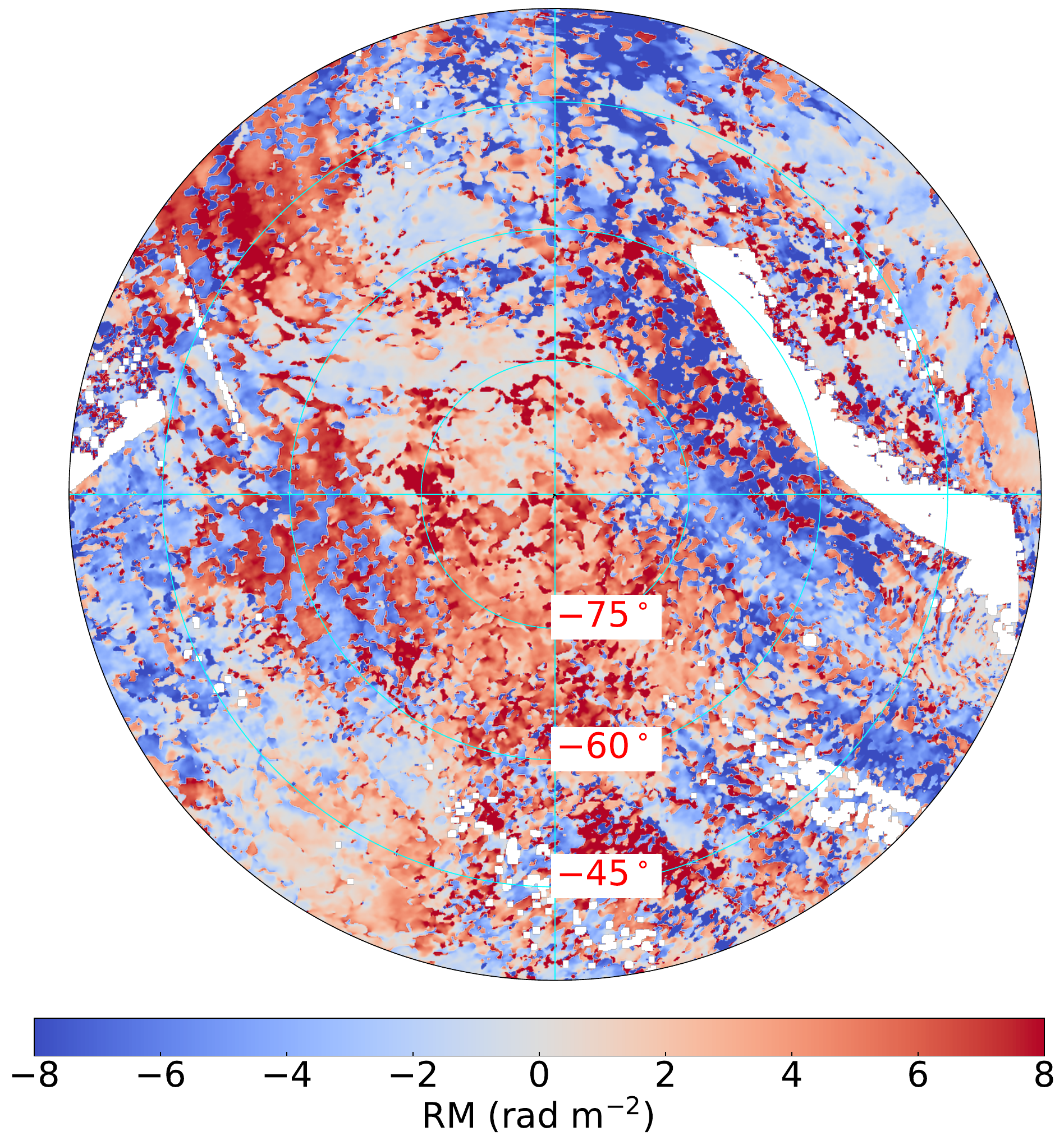}
    \caption{\it The same as Fig.~\ref{fig:pi_mwa} but for RM.}
    \label{fig:rm_mwa}
\end{figure}

For MWA, early observations covering 2400~deg$^2$ at 189~MHz with a 32-element prototype~\citep{Bernardi+13} revealed a wealth of diffuse polarized emission with RMs mainly positive and smaller than 10~rad~m$^{-2}$, at a maximum level of 13~K. Observations by \citet{Lenc+16} at 154~MHz found polarization structures of $2\degr-8\degr$ with an average brightness temperature of 4~K. The RM peaks at about 1~rad~m$^{-2}$, putting the structures at a distance of about 50~pc. \citet{Byrne+22} obtained band averaged $Q$ and $U$ images at 182~MHz and found that the emission structures were predominantly unpolarized, which could be caused by bandwidth depolarization. 

We reprocessed the highest frequency band (200-230~MHz) data from the GaLactic and Extragalactic All-sky MWA survey~\citep[GLEAM,][]{Wayth+15} following the same procedure as described by \citet{Lenc+16}. The off-axis leakage correction and the correction of Faraday rotation caused by ionosphere were done in the image plane. By tapering the baselines larger than $82\lambda$, the synthesized beam is very close to Gaussian with a width of about $1\degr$, and thus deconvolution is not needed. We divided the band into 24 channels each with a width of 1.28~MHz, and the FWHM of the RMSF is about 6~rad~m$^{-2}$. We imaged $Q$ and $U$ for each channel and then run RM synthesis to obtain the polarized intensity and RM images. Data processing is still being developed. The preliminary images covering a sky area of about 10,\,000~deg$^2$ (Sun et al. in prep.) are shown in Figs.~\ref{fig:pi_mwa} and \ref{fig:rm_mwa}. Areas with bad data or calibration are masked. Polarized structures of tens of degrees can be seen in these images. RM clearly exhibits longitude and latitude-dependent patterns, which is probably related to the LB.

\section{An SKA-Low polarization survey and scientific impact}

\subsection{A polarization survey}

The LOFAR observations of the diffuse polarized emission have high resolution but miss large-scale emission that is critical to interpret polarized structures. The MWA observations, as shown in Figs.~\ref{fig:pi_mwa} and \ref{fig:rm_mwa}, better represent large-scale emission, but at a coarser resolution.  

\begin{figure}[!htbp]
    \centering
	\includegraphics[width=0.98\columnwidth]{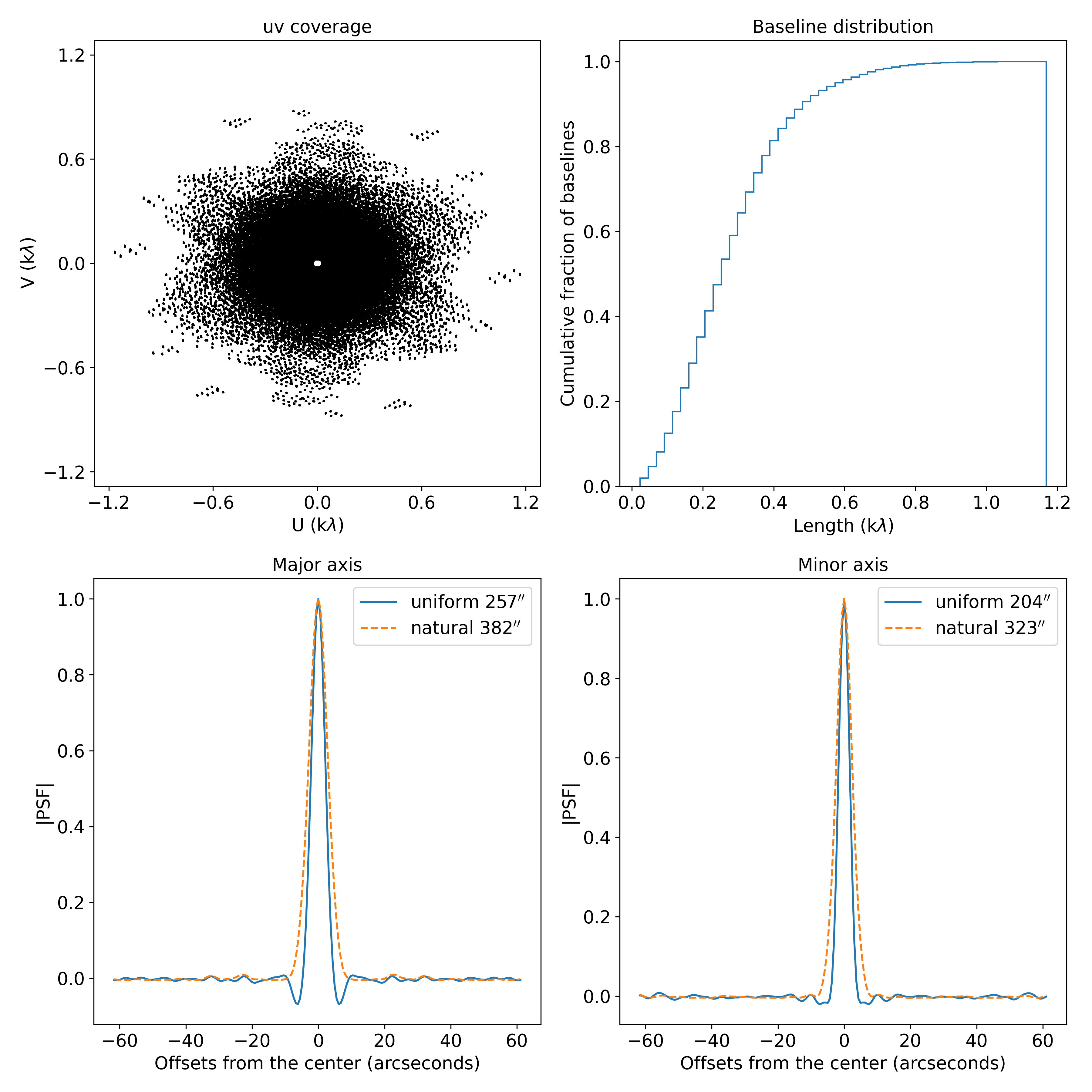}
    \caption{\it $uv$ coverage, baseline distribution, and sizes of the point spread function (PSF) for a 2~min snapshot observations of a source at $\delta=-60\degr$ at a single 1~MHz channel at 200~MHz, using stations within 1~km distance.}
    \label{fig:simu}
\end{figure}

A low-frequency survey by SKA-Low will improve the observations of diffuse emission. The SKA-Low is being built in stages (SKAO-TEL-0002299) with AA* containing 307 stations and AA4 containing 512 stations. Each station contains 256 log-periodic dipoles. 

We focus on the inner stations with distance less than 1~km, which provide high surface brightness sensitivity. We simulated a 2~min snapshot observation of a source at declination $\delta=-60\degr$ with a single 1~MHz frequency channel at 200~MHz, using the package \verb|ska_ost_array_config|\footnote{https://gitlab.com/ska-telescope/ost/ska-ost-array-config} provided by the SKA Observatory (SKAO). The $uv$ coverage, the distribution of the baselines, and the sizes of the point spread function (PSF) or the synthesized beam along the major and minor axes are shown in Fig.~\ref{fig:simu}. Compared to the $uv$ distribution of LoTSS~\citep[][their Fig.~2]{Shimwell+17}, SKA-Low provides a better coverage of short baselines. The PSF sizes from a uniform weighting are close to those from a natural weighting. The synthesized beam from natural weighting is symmetric and has very low levels of sidelobes, meaning that deconvolution is not needed. The resolution is about $6\arcmin$ for the natural weighting. The minimum baseline is about 0.02~k$\lambda$, corresponding to a maximum angular scale of about $3\degr$. Note that the simulation shown in Fig.~\ref{fig:simu} used the whole stations. The configuration of each SKA-Low station is very flexible, and substation arrangements can be utilized to incorporate baselines at least as short as those used with MWA to produce Figs.~\ref{fig:pi_mwa} and \ref{fig:rm_mwa}. If all the SKA-Low stations are used with substation configuration, unprecedented images of polarized emission with high resolution of arcseconds and large scales of degrees will be achieved, which will revolutionize the understanding of the local ISM.

SKA-Low covers the frequency range of 50-350~MHz. The difference in the Faraday rotation between the two ends of the frequency range is about a factor of 50. An RM causing complete depolarization at the low-frequency end could cause no depolarization at all at the high-frequency end. Therefore, observations at different frequencies detect polarized emission from different distances. We therefore suggest that the total frequency band be divided into subbands each with a width of 30~MHz. Each subband is further divided into narrow frequency channels to run RM synthesis. The channel width of 0.05~MHz for the lowest frequency subband and 0.1~MHz for the rest of the subbands are expected to be sufficient for $|{\rm RM}|$ less than around 10~rad~m$^{-2}$. The FWHM of the RMSF is approximately 1~rad~m$^{-2}$ at the low-frequency end.  

Sensitivity is difficult to assess. For total intensity, it probably reaches the confusion limit even for a 2~min observation. But the confusion limit in polarization is much lower. For the polarization images from LoTSS, the sensitivity is about 0.1~K. We expect a higher sensitivity from the SKA-Low polarization survey.

We advocate for an SKA-Low polarization survey to be conducted by AA*. The number of stations within 1~km distance is 215 for AA* and 260 for AA4. For surface brightness sensitivity, there will be an improvement of about 20\%. However, the timeline of AA4 is uncertain due to funding, and there will be a planned pause after AA*.

An SKA-Low polarization survey will produce images of polarized diffuse emission with high resolution and retaining large-scale structures. By applying RM synthesis, we will obtain precise RMs of these emission structures. 

\subsection{Scientific impact}

\subsubsection{The local insterstellar medium and local bubble}

The dense all-sky RM grid and the diffuse polarized emission from an SKA-Mid Band 2 polarization survey is well suited to reveal the Galactic magnetic field, but is not suitable for studying the LISM. It is difficult to separate the contribution of the LB to RM and emission from the rest of the Galactic ISM. In contrast, the LISM is the niche of the SKA-Low with a polarization survey. 

Images of diffuse polarized emission and RMs from an SKA-Low polarization survey will help to delineate the structure of the LB. As estimated earlier, an internal $|{\rm RM}|$ of 5~rad~m$^{-2}$ at 200~MHz already causes a nearly complete depolarization. However, RMs of the polarized structures from LOFAR and MWA span a range greater than $|{\rm RM}|$ of 5~rad~m$^{-2}$. This indicates that a large fraction of the polarized structures detected at low frequencies is Faraday thin, meaning that the internal RM is small and the RM comes mainly from the medium in front of the emission. With uniform electron density and magnetic field, RM is an indicator of the distance to the polarized emission. The diffuse polarized emission thus leads to a 3D structure of the LB. 

The current models of the LB assume a shell with an enhanced magnetic field, thermal and cosmic-ray electron densities. This means that polarized emission and RMs are contributed by the shell. In this case, the synchrotron-emitting and Faraday-rotating gasses are mixed, and strong wavelength-dependent depolarization is expected. The 3D structure of the shell can be constrained particularly with polarized emission at the low-frequency end. With the simulations by \citet{Maconi+25}, the model of the LB can be derived by fitting the observed RMs and polarized emission. With simulations~\citep[e.g.][]{Bracco+22}, RM contributions from the partially ionized warm and neutral medium in addition to the fully ionized gas can also be examined. 

Images of diffuse polarized emission also offer an opportunity to study turbulent magnetic fields using techniques such as gradient~\citep{Lazarian+24}. The path length is short, and thus the influence of line-of-sight integral can be largely circumvented. The scenario of the LB caused by sequential supernovae is expected to produce large turbulent fields, which can be investigated.  

The polarized emission with associated RMs provides a way to assess the foreground influence on observations of the epoch of reionization. In reverse to RM synthesis, mock observations of $\mathcal{P}(\lambda^2)$ can be derived from $F(\phi)$~\citep[e.g.][]{Spinelli+18}. In this way, the simulations are imprinted with the real structures of the Galactic ISM.  

The Discovering Sky at the Longest Wavelength (DSL) project~\citep[``Hongmeng" in Chinese,][]{Chen+23}, an interferometer formed with satellites orbiting the Moon, will be launched in the near future. DSL covers the frequency range of 0.1-30~MHz and will produce images of the sky at the longest wavelength. The polarization images together with those from SKA-Low will reveal the structure of ISM from local to very local. 

\subsubsection{Synergy between SKA-Low and SKA-Mid}

Low-frequency polarization data from LOFAR are revealing unexpected magnetic interactions within the multiphase ISM, where synchrotron polarization has been found to correlate with tracers of the partially ionized cold ISM, such as {\sc H\,i} and dust emission \citep{VanEck+17, Jelic+18, Bracco+20}, over large sky areas along the wall of the LB \citep{Erceg+22, Erceg+24}.

Faraday tomography is a powerful technique to probe the multiphase, magnetized gas. In the future, with the unprecedented wavelength coverage of the SKA, including both the SKA-Low and SKA-Mid instruments, it will be possible to overcome current limitations.

\begin{figure}[!htbp]
  \centering
  \includegraphics[width=0.98\columnwidth]{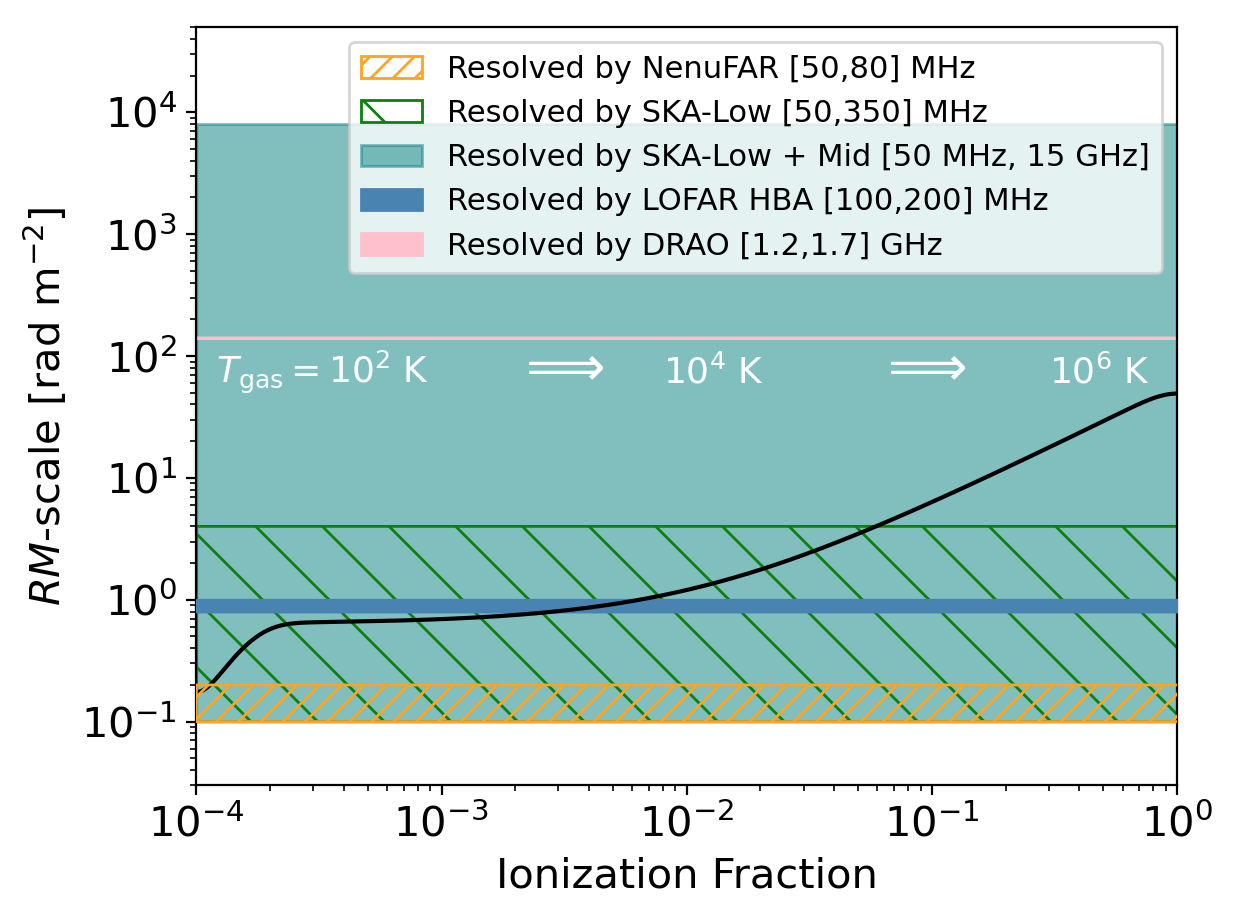} 
  \caption{\it RM-scale dependence on the ionization fraction in the multiphase ISM (in black), adapted from \citet{Ferriere20}. The gas temperature corresponding to the ISM phase is labeled in white. The RM-scale sensitivity of various facilities, including SKA, are shown in colors.}
\label{fig:rmxe}
\end{figure}

As in radio imaging, where the range of angular scales is set by the interferometric baseline coverage, a limited wavelength range restricts sensitivity to small and large scales in RM space. In Fig.~\ref{fig:rmxe}, we show in black the expected RM scales of different interstellar gas phases, represented by their ionization fraction and temperature \citep[adapted from][]{Ferriere20}, together with the RM-scale sensitivity of various telescopes—such as DRAO, LOFAR, and NenuFAR—ranging from high to low frequencies. These instruments strongly filter the RM space, resolving only restricted RM scales that correspond to different regimes of ionized gas. While high-frequency data allow us to explore fully ionized gas, low-frequency observations access RM scales associated with poorly and partially ionized phases on the verge of molecular cloud formation.

Only the SKA will fully populate this diagram, representing a breakthrough for complete multiscale studies of Faraday tomography. The combination of SKA-Low and SKA-Mid will potentially enable access to all available RM scales in the Galactic ISM, revealing detailed interactions between magnetic fields and multiphase interstellar gas in various regions, including the LB.

Even SKA-Low alone will represent a major leap forward, extending the accessible RM scales by nearly two orders of magnitude compared with existing facilities. This exceptional capability will not only provide invaluable insights into the magnetization of the local ISM, but, by tracing multiscale structures in poorly ionized gas, it will also enable commensal studies such as the investigation of diffuse CO-dark molecular gas---with ionization fraction of $\sim10^{-4}$---recently detected through formaldehyde spectroscopic observations combining centimetre and millimetre data \citep{Gerin+24}, thereby opening exciting prospects for synergies between SKA and the Atacama Large Millimeter Array (ALMA).

\section{Summary}

We are located inside the LB, the origin and evolution of which have not yet been solved. It has been shown that the LB makes a considerable contribution to dust polarization, RM, and synchrotron polarized emission. 
The LISM including the LB is a niche for SKA-Low with an polarization survey. At the frequency range of 50-350~MHz, the polarized emission is predominantly from a distance less than about 300~pc, probably related to the LB. An SKA-Low polarization survey will produce images of diffuse polarized emission and precise RMs. This will help to reveal the 3D structures of the LB and thus understand how the LB forms and evolves.

\section*{Acknowledgments}
XS is supported by the National SKA Program of China (Grant No. 2022SKA0120101) and the National Natural Science Foundation of China (No. 12433006). AB acknowledges financial support from the INAF initiative ``IAF Astronomy Fellowships in Italy'' (grant name MEGASKAT).

\bibliographystyle{abbrvnat-maxbibnames4}
\bibliography{chapter}

\begin{thebibliography}{47}
\providecommand{\natexlab}[1]{#1}
\providecommand{\url}[1]{\texttt{#1}}
\expandafter\ifx\csname urlstyle\endcsname\relax
  \providecommand{\doi}[1]{doi: #1}\else
  \providecommand{\doi}{doi: \begingroup \urlstyle{rm}\Url}\fi

\bibitem[{Alves} et~al.(2018){Alves}, {Boulanger}, {Ferri{\`e}re}, and {Montier}]{Alves+18}
M.~I.~R. {Alves}, F.~{Boulanger}, K.~{Ferri{\`e}re}, and L.~{Montier}.
\newblock \emph{\aap}, 611:\penalty0 L5, Mar. 2018.
\newblock \doi{10.1051/0004-6361/201832637}.

\bibitem[{Bernardi} et~al.(2013){Bernardi}, {Greenhill}, {Mitchell}, {Ord}, {Hazelton}, {Gaensler}, {de Oliveira-Costa}, {Morales}, {Udaya Shankar}, {Subrahmanyan}, {Wayth}, {Lenc}, {Williams}, {Arcus}, {Arora}, {Barnes}, {Bowman}, {Briggs}, {Bunton}, {Cappallo}, {Corey}, {Deshpande}, {deSouza}, {Emrich}, {Goeke}, {Herne}, {Hewitt}, {Johnston-Hollitt}, {Kaplan}, {Kasper}, {Kincaid}, {Koenig}, {Kratzenberg}, {Lonsdale}, {Lynch}, {McWhirter}, {Morgan}, {Oberoi}, {Pathikulangara}, {Prabu}, {Remillard}, {Rogers}, {Roshi}, {Salah}, {Sault}, {Srivani}, {Stevens}, {Tingay}, {Waterson}, {Webster}, {Whitney}, {Williams}, and {Wyithe}]{Bernardi+13}
G.~{Bernardi} et al.
\newblock \emph{\apj}, 771\penalty0 (2):\penalty0 105, July 2013.
\newblock \doi{10.1088/0004-637X/771/2/105}.

\bibitem[{Boulanger} et~al.(2024){Boulanger}, {Gry}, {Jenkins}, {Bracco}, {Erceg}, {Jeli{\'c}}, and {Turi{\'c}}]{Boulanger+24}
F.~{Boulanger} et al.
\newblock \emph{\aap}, 687:\penalty0 A102, July 2024.
\newblock \doi{10.1051/0004-6361/202348953}.

\bibitem[{Bracco} et~al.(2020){Bracco}, {Jeli{\'c}}, {Marchal}, {Turi{\'c}}, {Erceg}, {Miville-Desch{\^e}nes}, and {Bellomi}]{Bracco+20}
A.~{Bracco} et al.
\newblock \emph{\aap}, 644:\penalty0 L3, Dec. 2020.
\newblock \doi{10.1051/0004-6361/202039283}.

\bibitem[{Bracco} et~al.(2022){Bracco}, {Ntormousi}, {Jeli{\'c}}, {Padovani}, {{\v{S}}iljeg}, {Erceg}, {Turi{\'c}}, {Ceraj}, and {{\v{S}}nidari{\'c}}]{Bracco+22}
A.~{Bracco} et al.
\newblock \emph{\aap}, 663:\penalty0 A37, July 2022.
\newblock \doi{10.1051/0004-6361/202142453}.

\bibitem[{Breitschwerdt} et~al.(2016){Breitschwerdt}, {Feige}, {Schulreich}, {Avillez}, {Dettbarn}, and {Fuchs}]{Breitschwerdt+16}
D.~{Breitschwerdt} et al.
\newblock \emph{\nat}, 532\penalty0 (7597):\penalty0 73--76, Apr. 2016.
\newblock \doi{10.1038/nature17424}.

\bibitem[{Brentjens} and {de Bruyn}(2005)]{Brentjens+05}
M.~A. {Brentjens} and A.~G. {de Bruyn}.
\newblock \emph{\aap}, 441\penalty0 (3):\penalty0 1217--1228, Oct. 2005.
\newblock \doi{10.1051/0004-6361:20052990}.

\bibitem[{Burn}(1966)]{Burn+66}
B.~J. {Burn}.
\newblock \emph{\mnras}, 133:\penalty0 67, Jan. 1966.
\newblock \doi{10.1093/mnras/133.1.67}.

\bibitem[{Byrne} et~al.(2022){Byrne}, {Morales}, {Hazelton}, {Sullivan}, {Barry}, {Lynch}, {Line}, and {Jacobs}]{Byrne+22}
R.~{Byrne} et al.
\newblock \emph{\mnras}, 510\penalty0 (2):\penalty0 2011--2024, Feb. 2022.
\newblock \doi{10.1093/mnras/stab3276}.

\bibitem[{Chen} et~al.(2023){Chen}, {Yan}, {Xu}, {Deng}, {Wu}, {Wu}, {Zhou}, {Zhang}, {Zhu}, {Yang}, and {Wu}]{Chen+23}
X.~{Chen} et al.
\newblock \emph{Chinese Journal of Space Science}, 43\penalty0 (1):\penalty0 43--59, Jan. 2023.
\newblock \doi{10.11728/cjss2023.01.220104001}.

\bibitem[{Cox} and {Reynolds}(1987)]{Cox+87}
D.~P. {Cox} and R.~J. {Reynolds}.
\newblock \emph{\araa}, 25:\penalty0 303--344, Jan. 1987.
\newblock \doi{10.1146/annurev.aa.25.090187.001511}.

\bibitem[{Edenhofer} et~al.(2024){Edenhofer}, {Zucker}, {Frank}, {Saydjari}, {Speagle}, {Finkbeiner}, and {En{\ss}lin}]{Edenhofer+24}
G.~{Edenhofer} et al.
\newblock \emph{\aap}, 685:\penalty0 A82, May 2024.
\newblock \doi{10.1051/0004-6361/202347628}.

\bibitem[{Erceg} et~al.(2022){Erceg}, {Jeli{\'c}}, {Haverkorn}, {Bracco}, {Shimwell}, {Tasse}, {Dickey}, {Ceraj}, {Drabent}, {Hardcastle}, and {Turi{\'c}}]{Erceg+22}
A.~{Erceg} et al.
\newblock \emph{\aap}, 663:\penalty0 A7, July 2022.
\newblock \doi{10.1051/0004-6361/202142244}.

\bibitem[{Erceg} et~al.(2024{\natexlab{a}}){Erceg}, {Jeli{\'c}}, {Haverkorn}, {Bracco}, {Ceraj}, {Turi{\'c}}, and {Soler}]{Erceg+24a}
A.~{Erceg} et al.
\newblock \emph{\aap}, 687:\penalty0 A23, July 2024{\natexlab{a}}.
\newblock \doi{10.1051/0004-6361/202348586}.

\bibitem[{Erceg} et~al.(2024{\natexlab{b}}){Erceg}, {Jeli{\'c}}, {Haverkorn}, {Gajovi{\'c}}, {Hardcastle}, {Shimwell}, and {Tasse}]{Erceg+24}
A.~{Erceg} et al.
\newblock \emph{\aap}, 688:\penalty0 A200, Aug. 2024{\natexlab{b}}.
\newblock \doi{10.1051/0004-6361/202450082}.

\bibitem[{Ferri{\`e}re}(2020)]{Ferriere20}
K.~{Ferri{\`e}re}.
\newblock \emph{Plasma Physics and Controlled Fusion}, 62\penalty0 (1):\penalty0 014014, Jan. 2020.
\newblock \doi{10.1088/1361-6587/ab49eb}.

\bibitem[{Fuchs} et~al.(2006){Fuchs}, {Breitschwerdt}, {de Avillez}, {Dettbarn}, and {Flynn}]{Fuchs+06}
B.~{Fuchs} et al.
\newblock \emph{\mnras}, 373\penalty0 (3):\penalty0 993--1003, Dec. 2006.
\newblock \doi{10.1111/j.1365-2966.2006.11044.x}.

\bibitem[{Gaensler} et~al.(2001){Gaensler}, {Dickey}, {McClure-Griffiths}, {Green}, {Wieringa}, and {Haynes}]{Gaensler+01}
B.~M. {Gaensler} et al.
\newblock \emph{\apj}, 549\penalty0 (2):\penalty0 959--978, Mar. 2001.
\newblock \doi{10.1086/319468}.

\bibitem[{Gerin} et~al.(2024){Gerin}, {Liszt}, {Pety}, and {Faure}]{Gerin+24}
M.~{Gerin}, H.~{Liszt}, J.~{Pety}, and A.~{Faure}.
\newblock \emph{\aap}, 686:\penalty0 A49, June 2024.
\newblock \doi{10.1051/0004-6361/202449152}.

\bibitem[{Han}(2017)]{Han+17}
J.~L. {Han}.
\newblock \emph{\araa}, 55\penalty0 (1):\penalty0 111--157, Aug. 2017.
\newblock \doi{10.1146/annurev-astro-091916-055221}.

\bibitem[{Haverkorn}(2015)]{Haverkorn+15}
M.~{Haverkorn}.
\newblock In A.~{Lazarian}, E.~M. {de Gouveia Dal Pino}, and C.~{Melioli}, editors, \emph{Magnetic Fields in Diffuse Media}, volume 407 of \emph{Astrophysics and Space Science Library}, page 483, Jan. 2015.
\newblock \doi{10.1007/978-3-662-44625-6_17}.

\bibitem[{Jeli{\'c}} et~al.(2014){Jeli{\'c}}, {de Bruyn}, {Mevius}, {Abdalla}, {Asad}, {Bernardi}, {Brentjens}, {Bus}, {Chapman}, {Ciardi}, {Daiboo}, {Fernandez}, {Ghosh}, {Harker}, {Jensen}, {Kazemi}, {Koopmans}, {Labropoulos}, {Martinez-Rubi}, {Mellema}, {Offringa}, {Pandey}, {Patil}, {Thomas}, {Vedantham}, {Veligatla}, {Yatawatta}, {Zaroubi}, {Alexov}, {Anderson}, {Avruch}, {Beck}, {Bell}, {Bentum}, {Best}, {Bonafede}, {Bregman}, {Breitling}, {Broderick}, {Brouw}, {Br{\"u}ggen}, {Butcher}, {Conway}, {de Gasperin}, {de Geus}, {Deller}, {Dettmar}, {Duscha}, {Eisl{\"o}ffel}, {Engels}, {Falcke}, {Fallows}, {Fender}, {Ferrari}, {Frieswijk}, {Garrett}, {Grie{\ss}meier}, {Gunst}, {Hamaker}, {Hassall}, {Haverkorn}, {Heald}, {Hessels}, {Hoeft}, {H{\"o}randel}, {Horneffer}, {van der Horst}, {Iacobelli}, {Juette}, {Karastergiou}, {Kondratiev}, {Kramer}, {Kuniyoshi}, {Kuper}, {van Leeuwen}, {Maat}, {Mann}, {McKay-Bukowski}, {McKean}, {Munk}, {Nelles}, {Norden}, {Paas}, {Pandey-Pommier}, {Pietka}, {Pizzo}, {Polatidis},
  {Reich}, {R{\"o}ttgering}, {Rowlinson}, {Scaife}, {Schwarz}, {Serylak}, {Smirnov}, {Steinmetz}, {Stewart}, {Tagger}, {Tang}, {Tasse}, {ter Veen}, {Thoudam}, {Toribio}, {Vermeulen}, {Vocks}, {van Weeren}, {Wijers}, {Wijnholds}, {Wucknitz}, and {Zarka}]{Jelic+14}
V.~{Jeli{\'c}} et al.
\newblock \emph{\aap}, 568:\penalty0 A101, Aug. 2014.
\newblock \doi{10.1051/0004-6361/201423998}.

\bibitem[{Jeli{\'c}} et~al.(2018){Jeli{\'c}}, {Prelogovi{\'c}}, {Haverkorn}, {Remeijn}, and {Klind{\v{z}}i{\'c}}]{Jelic+18}
V.~{Jeli{\'c}} et al.
\newblock \emph{\aap}, 615:\penalty0 L3, July 2018.
\newblock \doi{10.1051/0004-6361/201833291}.

\bibitem[{Korochkin} et~al.(2025){Korochkin}, {Semikoz}, and {Tinyakov}]{Korochkin+25}
A.~{Korochkin}, D.~{Semikoz}, and P.~{Tinyakov}.
\newblock \emph{\aap}, 693:\penalty0 A284, Jan. 2025.
\newblock \doi{10.1051/0004-6361/202451440}.

\bibitem[{Lazarian} et~al.(2024){Lazarian}, {Yuen}, and {Pogosyan}]{Lazarian+24}
A.~{Lazarian}, K.~H. {Yuen}, and D.~{Pogosyan}.
\newblock \emph{\apj}, 974\penalty0 (2):\penalty0 237, Oct. 2024.
\newblock \doi{10.3847/1538-4357/ad6d62}.

\bibitem[{Lenc} et~al.(2016){Lenc}, {Gaensler}, {Sun}, {Sadler}, {Willis}, {Barry}, {Beardsley}, {Bell}, {Bernardi}, {Bowman}, {Briggs}, {Callingham}, {Cappallo}, {Carroll}, {Corey}, {de Oliveira-Costa}, {Deshpande}, {Dillon}, {Dwarkanath}, {Emrich}, {Ewall-Wice}, {Feng}, {For}, {Goeke}, {Greenhill}, {Hancock}, {Hazelton}, {Hewitt}, {Hindson}, {Hurley-Walker}, {Johnston-Hollitt}, {Jacobs}, {Kapi{\'n}ska}, {Kaplan}, {Kasper}, {Kim}, {Kratzenberg}, {Line}, {Loeb}, {Lonsdale}, {Lynch}, {McKinley}, {McWhirter}, {Mitchell}, {Morales}, {Morgan}, {Morgan}, {Murphy}, {Neben}, {Oberoi}, {Offringa}, {Ord}, {Paul}, {Pindor}, {Pober}, {Prabu}, {Procopio}, {Riding}, {Rogers}, {Roshi}, {Udaya Shankar}, {Sethi}, {Srivani}, {Staveley-Smith}, {Subrahmanyan}, {Sullivan}, {Tegmark}, {Thyagarajan}, {Tingay}, {Trott}, {Waterson}, {Wayth}, {Webster}, {Whitney}, {Williams}, {Williams}, {Wu}, {Wyithe}, and {Zheng}]{Lenc+16}
E.~{Lenc} et al.
\newblock \emph{\apj}, 830\penalty0 (1):\penalty0 38, Oct. 2016.
\newblock \doi{10.3847/0004-637X/830/1/38}.

\bibitem[{Maconi} et~al.(2023){Maconi}, {Soler}, {Reissl}, {Girichidis}, {Klessen}, {Hennebelle}, {Molinari}, {Testi}, {Smith}, {Sormani}, {Teh}, and {Traficante}]{Maconi+23}
E.~{Maconi} et al.
\newblock \emph{\mnras}, 523\penalty0 (4):\penalty0 5995--6010, Aug. 2023.
\newblock \doi{10.1093/mnras/stad1854}.

\bibitem[{Maconi} et~al.(2025){Maconi}, {Reissl}, {Soler}, {Girichidis}, {Klessen}, {Bracco}, and {Hutschenreuter}]{Maconi+25}
E.~{Maconi} et al.
\newblock \emph{\aap}, 698:\penalty0 A84, June 2025.
\newblock \doi{10.1051/0004-6361/202451477}.

\bibitem[{Nord} et~al.(2006){Nord}, {Henning}, {Rand}, {Lazio}, and {Kassim}]{Nord+06}
M.~E. {Nord} et al.
\newblock \emph{\aj}, 132\penalty0 (1):\penalty0 242--252, July 2006.
\newblock \doi{10.1086/504407}.

\bibitem[{O'Neill} et~al.(2024){O'Neill}, {Zucker}, {Goodman}, and {Edenhofer}]{O'Neill+24}
T.~J. {O'Neill}, C.~{Zucker}, A.~A. {Goodman}, and G.~{Edenhofer}.
\newblock \emph{\apj}, 973\penalty0 (2):\penalty0 136, Oct. 2024.
\newblock \doi{10.3847/1538-4357/ad61de}.

\bibitem[{O'Neill} et~al.(2025){O'Neill}, {Goodman}, {Soler}, {Zucker}, and {Han}]{O'Neill+25}
T.~J. {O'Neill} et al.
\newblock \emph{\apj}, 988\penalty0 (2):\penalty0 191, Aug. 2025.
\newblock \doi{10.3847/1538-4357/ade306}.

\bibitem[{Pelgrims} et~al.(2025){Pelgrims}, {Unger}, and {Mari{\c{s}}}]{Pelgrims+25}
V.~{Pelgrims}, M.~{Unger}, and I.~C. {Mari{\c{s}}}.
\newblock \emph{\aap}, 695:\penalty0 A148, Mar. 2025.
\newblock \doi{10.1051/0004-6361/202452943}.

\bibitem[{Roger} et~al.(1999){Roger}, {Costain}, {Landecker}, and {Swerdlyk}]{Roger+99}
R.~S. {Roger}, C.~H. {Costain}, T.~L. {Landecker}, and C.~M. {Swerdlyk}.
\newblock \emph{\aaps}, 137:\penalty0 7--19, May 1999.
\newblock \doi{10.1051/aas:1999239}.

\bibitem[{Shimwell} et~al.(2017){Shimwell}, {R{\"o}ttgering}, {Best}, {Williams}, {Dijkema}, {de Gasperin}, {Hardcastle}, {Heald}, {Hoang}, {Horneffer}, {Intema}, {Mahony}, {Mandal}, {Mechev}, {Morabito}, {Oonk}, {Rafferty}, {Retana-Montenegro}, {Sabater}, {Tasse}, {van Weeren}, {Br{\"u}ggen}, {Brunetti}, {Chy{\.z}y}, {Conway}, {Haverkorn}, {Jackson}, {Jarvis}, {McKean}, {Miley}, {Morganti}, {White}, {Wise}, {van Bemmel}, {Beck}, {Brienza}, {Bonafede}, {Calistro Rivera}, {Cassano}, {Clarke}, {Cseh}, {Deller}, {Drabent}, {van Driel}, {Engels}, {Falcke}, {Ferrari}, {Fr{\"o}hlich}, {Garrett}, {Harwood}, {Heesen}, {Hoeft}, {Horellou}, {Israel}, {Kapi{\'n}ska}, {Kunert-Bajraszewska}, {McKay}, {Mohan}, {Orr{\'u}}, {Pizzo}, {Prandoni}, {Schwarz}, {Shulevski}, {Sipior}, {Smith}, {Sridhar}, {Steinmetz}, {Stroe}, {Varenius}, {van der Werf}, {Zensus}, and {Zwart}]{Shimwell+17}
T.~W. {Shimwell} et al.
\newblock \emph{\aap}, 598:\penalty0 A104, Feb. 2017.
\newblock \doi{10.1051/0004-6361/201629313}.

\bibitem[{Sokoloff} et~al.(1998){Sokoloff}, {Bykov}, {Shukurov}, {Berkhuijsen}, {Beck}, and {Poezd}]{Sokoloff+98}
D.~D. {Sokoloff} et al.
\newblock \emph{\mnras}, 299\penalty0 (1):\penalty0 189--206, Aug. 1998.
\newblock \doi{10.1046/j.1365-8711.1998.01782.x}.

\bibitem[{Spinelli} et~al.(2018){Spinelli}, {Bernardi}, and {Santos}]{Spinelli+18}
M.~{Spinelli}, G.~{Bernardi}, and M.~G. {Santos}.
\newblock \emph{\mnras}, 479\penalty0 (1):\penalty0 275--283, Sept. 2018.
\newblock \doi{10.1093/mnras/sty1457}.

\bibitem[{Sun} and {Reich}(2010)]{sun+10}
X.-H. {Sun} and W.~{Reich}.
\newblock \emph{Research in Astronomy and Astrophysics}, 10\penalty0 (12):\penalty0 1287--1297, Dec. 2010.
\newblock \doi{10.1088/1674-4527/10/12/009}.

\bibitem[{Sun} et~al.(2008){Sun}, {Reich}, {Waelkens}, and {En{\ss}lin}]{Sun+08}
X.~H. {Sun}, W.~{Reich}, A.~{Waelkens}, and T.~A. {En{\ss}lin}.
\newblock \emph{\aap}, 477\penalty0 (2):\penalty0 573--592, Jan. 2008.
\newblock \doi{10.1051/0004-6361:20078671}.

\bibitem[{Sun} et~al.(2015){Sun}, {Rudnick}, {Akahori}, {Anderson}, {Bell}, {Bray}, {Farnes}, {Ideguchi}, {Kumazaki}, {O'Brien}, {O'Sullivan}, {Scaife}, {Stepanov}, {Stil}, {Takahashi}, {van Weeren}, and {Wolleben}]{Sun+15}
X.~H. {Sun} et al.
\newblock \emph{\aj}, 149\penalty0 (2):\penalty0 60, Feb. 2015.
\newblock \doi{10.1088/0004-6256/149/2/60}.

\bibitem[{Turi{\'c}} et~al.(2021){Turi{\'c}}, {Jeli{\'c}}, {Jaspers}, {Haverkorn}, {Bracco}, {Erceg}, {Ceraj}, {van Eck}, and {Zaroubi}]{Turic+21}
L.~{Turi{\'c}} et al.
\newblock \emph{\aap}, 654:\penalty0 A5, Oct. 2021.
\newblock \doi{10.1051/0004-6361/202141071}.

\bibitem[{Van Eck} et~al.(2017){Van Eck}, {Haverkorn}, {Alves}, {Beck}, {de Bruyn}, {En{\ss}lin}, {Farnes}, {Ferri{\`e}re}, {Heald}, {Horellou}, {Horneffer}, {Iacobelli}, {Jeli{\'c}}, {Mart{\'\i}-Vidal}, {Mulcahy}, {Reich}, {R{\"o}ttgering}, {Scaife}, {Schnitzeler}, {Sobey}, and {Sridhar}]{VanEck+17}
C.~L. {Van Eck} et al.
\newblock \emph{\aap}, 597:\penalty0 A98, Jan. 2017.
\newblock \doi{10.1051/0004-6361/201629707}.

\bibitem[{Van Eck} et~al.(2019){Van Eck}, {Haverkorn}, {Alves}, {Beck}, {Best}, {Carretti}, {Chy{\.z}y}, {En{\ss}lin}, {Farnes}, {Ferri{\`e}re}, {Heald}, {Iacobelli}, {Jeli{\'c}}, {Reich}, {R{\"o}ttgering}, and {Schnitzeler}]{VanEck+19}
C.~L. {Van Eck} et al.
\newblock \emph{\aap}, 623:\penalty0 A71, Mar. 2019.
\newblock \doi{10.1051/0004-6361/201834777}.

\bibitem[{Vanderwoude} et~al.(2024){Vanderwoude}, {West}, {Gaensler}, {Rudnick}, {Van Eck}, {Thomson}, {Andernach}, {Anderson}, {Carretti}, {Heald}, {Leahy}, {McClure-Griffiths}, {O'Sullivan}, {Tahani}, and {Willis}]{Vanderwoude+24}
S.~{Vanderwoude} et al.
\newblock \emph{\aj}, 167\penalty0 (5):\penalty0 226, May 2024.
\newblock \doi{10.3847/1538-3881/ad2fc8}.

\bibitem[{{\v{S}}nidari{\'c}} et~al.(2023){{\v{S}}nidari{\'c}}, {Jeli{\'c}}, {Mevius}, {Brentjens}, {Erceg}, {Shimwell}, {Piras}, {Horellou}, {Sabater}, {Best}, {Bracco}, {Ceraj}, {Haverkorn}, {O'Sullivan}, {Turi{\'c}}, and {Vacca}]{Snidaric+23}
I.~{{\v{S}}nidari{\'c}} et al.
\newblock \emph{\aap}, 674:\penalty0 A119, June 2023.
\newblock \doi{10.1051/0004-6361/202245124}.

\bibitem[{Wayth} et~al.(2015){Wayth}, {Lenc}, {Bell}, {Callingham}, {Dwarakanath}, {Franzen}, {For}, {Gaensler}, {Hancock}, {Hindson}, {Hurley-Walker}, {Jackson}, {Johnston-Hollitt}, {Kapi{\'n}ska}, {McKinley}, {Morgan}, {Offringa}, {Procopio}, {Staveley-Smith}, {Wu}, {Zheng}, {Trott}, {Bernardi}, {Bowman}, {Briggs}, {Cappallo}, {Corey}, {Deshpande}, {Emrich}, {Goeke}, {Greenhill}, {Hazelton}, {Kaplan}, {Kasper}, {Kratzenberg}, {Lonsdale}, {Lynch}, {McWhirter}, {Mitchell}, {Morales}, {Morgan}, {Oberoi}, {Ord}, {Prabu}, {Rogers}, {Roshi}, {Shankar}, {Srivani}, {Subrahmanyan}, {Tingay}, {Waterson}, {Webster}, {Whitney}, {Williams}, and {Williams}]{Wayth+15}
R.~B. {Wayth} et al.
\newblock \emph{\pasa}, 32:\penalty0 e025, June 2015.
\newblock \doi{10.1017/pasa.2015.26}.

\bibitem[{Yao} et~al.(2017){Yao}, {Manchester}, and {Wang}]{Yao+17}
J.~M. {Yao}, R.~N. {Manchester}, and N.~{Wang}.
\newblock \emph{\apj}, 835\penalty0 (1):\penalty0 29, Jan. 2017.
\newblock \doi{10.3847/1538-4357/835/1/29}.

\bibitem[{Zucker} et~al.(2022){Zucker}, {Goodman}, {Alves}, {Bialy}, {Foley}, {Speagle}, {Gro{\^I}{\texttwosuperior}schedl}, {Finkbeiner}, {Burkert}, {Khimey}, and {Swiggum}]{Zucker+22}
C.~{Zucker} et al.
\newblock \emph{\nat}, 601\penalty0 (7893):\penalty0 334--337, Jan. 2022.
\newblock \doi{10.1038/s41586-021-04286-5}.

\end{thebibliography}
\end{document}